\documentstyle[12pt,epsfig,amsmath,amssymb,psfrag,pifont,cite]{article}
\numberwithin{equation}{section}
\numberwithin{table}{section}
\newcommand{\lambdabf}{{\mbox{\boldmath $\lambda$}}}

\renewcommand{\d}{\partial}

\def\ga{\mathrel{\raise.3ex\hbox{$>$\kern-.75em\lower1ex\hbox{$\sim$}}}}
\def\la{\mathrel{\raise.3ex\hbox{$<$\kern-.75em\lower1ex\hbox{$\sim$}}}}

\def\I_M{{I_{\scriptscriptstyle M\times M}}}

\def\be{\begin{equation}}
\def\ee{\end{equation}}
\def\bea{\begin{eqnarray}}
\def\eea{\end{eqnarray}}

\newcommand{\beqal}{\begin{eqnarray}\label}
\newcommand{\beqa}{\begin{eqnarray}}
\newcommand{\eeqa}{\end{eqnarray}}

\newcommand {\f}{\frac}

\begin{document}

\begin{titlepage}
\begin{center}

\vskip .2in

{\Large \bf \textbf{Proof of universality of electrical conductivity at finite chemical potential}}
\vskip .5in

{ 
\bf Sayan K. Chakrabarti\footnote{e-mail: sayan@iopb.res.in}, \bf Shankhadeep Chakrabortty\footnote{e-mail: sankha@iopb.res.in},  \bf Sachin Jain\footnote{e-mail: sachjain@iopb.res.in}\\
\vskip .1in
{\em Institute of Physics,\\
Sachivalaya Marg, Bhubaneswar,\\ India-751~005.}}
\end{center}\noindent
\baselineskip 15pt

\begin{center} {\bf ABSTRACT}

\end{center}
\begin{quotation}\noindent
\baselineskip 15pt
It was proposed in \cite{Jain:2010ip} that, for certain gauge theories with gravity duals, electrical conductivity at finite chemical potential is universal. 
Here we provide a general proof that, when matter stress tensor satisfies a compact constraint, electrical conductivity is universal. We further elaborate our 
result with several conformal as well as non-conformal gauge theories. We also discuss how boundary conductivity and universal conductivity of stretched horizon are related.
\end{quotation}
\vskip 2in
\end{titlepage}
\vfill
\eject

\setcounter{footnote}{1}
\section{Introduction}

The fluid/gravity correspondence provides us with two distinct fluids dual to a given black hole geometry: first, the fluid given by membrane paradigm, which is described by quantities at the black hole horizon and second, the fluid at the boundary of the space time known from gauge/gravity duality and is described by quantities at the boundary. By exploiting the fact that changing radial position in the bulk corresponds to RG flow in the boundary fluid,  \cite{Iqbal:2008by,Bredberg:2010ky}
proposed a number of relations and even interpolation between them. For example, radial independence of certain quantities is used to show that, the shear viscosity ($\eta$) to entropy density ($s$) ratio ($\frac{\eta}{s}$) for both the fluid is same as well as the low frequency limit of  electrical conductivities of these two distinct fluids computed at zero chemical potential, are related. However, the situation changes significantly at finite chemical potential in the boundary theory (which corresponds to charged black hole in the bulk), where radial independence exploited earlier in relating electrical conductivity of these two fluids, gets completely destroyed.  One needs to solve flow equation in order to relate conductivities of these two fluids. Recently it was proposed in \cite{Jain:2010ip}, based on few 
examples that, the boundary electrical conductivity is universal and that there exists a simple relation between the conductivities of the fluids at horizon and at boundary. It was further proposed that, at any radial position $r$, the conductivity is given by a simple expression which interpolates smoothly between the one computed at the horizon and at the boundary. In all the examples considered in the above mentioned work, the bulk theory was asymptotically AdS for which there exists a CFT on the boundary at finite temperature and chemical potential. Now, going beyond CFT, it would be interesting to see whether such universality holds good for other boundary theories such as  non-conformal or Lifshitz like theories with finite chemical potentials. For gauge theories dual to charged Lifshitz like gravity backgrounds, it was shown in \cite{Jain:2010ip} that,
the above mentioned universality does not hold. On the contrary, the electrical conductivity of other theories such as non-conformal fluid living on charged  $D1$ brane \cite{David:2010qc} shows the same universality mentioned above (as shown later in the paper). So at present, there is no systematic way to answer which theories would show the universality in the electrical conductivity and which will not \footnote{Note that a closed formula for the electrical conductivity at finite chemical potential has been presented for a wide class of models in \cite{Mas:2008qs}. Although the set up we will consider in this paper will be completely different from the one used in \cite{Mas:2008qs} which involves probe branes.}. Rather than checking case by case, it is desirable to have a characterization for the theories which will show the proposed universality. In this paper, we show that only those gauge theories which  have a gravity dual with particular matter 
content will show this universality. This further explains why charged Lifshitz like black holes do not show the universality.

The paper is structured as follows. Section $2$ is a review of earlier works \cite{Jain:2010ip, Jain:2009pw, Myers:2009ij}.
In this section we also discuss all the assumptions made in the gravity side. In section $3$, we find the condition in the gravity
side under which the dual gauge theory will show the universality. Section $4$ discusses several examples, 
which include theories at and away from conformality. This section also explains why the Lifshitz like theories do not 
show the universality. In section $5$, we explicitly discuss on non-conformal gauge theory dual to charged $Dp$ brane,
and show the universality. Finally for completeness of this work we explicitly check  the universality for cases with multiple charges.  We further compute the thermal
 conductivity to viscosity ratio and show its universality. Finally we conclude the paper in section 6. In appendix $A$, we briefly discuss computation of electrical conductivity from gravity side and  discuss the flow equations. In appendix $B,$ we elaborate upon the condition that we get on energy momentum tensor.

\section{What to prove?}

This section is essentially a review of the earlier works. In this section we discuss the notations and assumptions in the gravity theory which are assumed to support 
gauge/gravity duality and write down the perturbation equation required for the computation of electrical conductivity (for details see \cite{Jain:2010ip}). Since we are 
interested in calculating the electrical conductivity in the presence of chemical potential, we consider the most general two-derivative gravity action of the 
following form 
\begin{eqnarray}
S =\frac{1}{2 \kappa^{2}}\int d^{d+1}x~~\sqrt{-g}(
R  -\frac{1}{4 g_{\rm{eff}}^2(r)} F_{\mu\nu}
 F^{\mu \nu} + \rm{Other~ terms}),
%\int d^{d+1}x~~\sqrt{-g}(\frac{1}{2 \kappa^{2}} 
%R  - \frac{1}{4 g_{d+1}^{2}} \widehat{G}(r) F_{\mu\nu}
 %F^{\mu \nu} + \rm{Other~ terms}),\nonumber\\
\label{lagrangian}
\end{eqnarray}
where $ F^{\mu \nu}$ is the field-strength tensor of the  $U(1)$ gauge field and $\frac{1}{g_{\rm{eff}}^2}$ is the 
effective gauge coupling. The metric that we take is of the form
\begin{equation}
 ds^{2} = g_{tt}(r) dt^{2} + g_{rr}(r) dr^{2} + g_{xx}(r) \sum_{i=1}^{d-1} (dx^{i})^{2},\label{met1}
\end{equation} 
where $r$ is the radial coordinate. We have assumed full rotational symmetry in $x^{i}$ directions so that\footnote{Let us note that, we are using the notation where $g_{\mu\nu}(r) \equiv g_{\mu\nu},$ $\frac{1}{g_{\rm{eff}}^2(r)}\equiv \frac{1}{g_{\rm{eff}}^2}.$} $g_{ij} = g_{xx} \delta_{ij}$, where $i,j$ run over all the indices except $r, t$. We also assume that metric components depend on radial coordinate only. We shall work with the metric which has an event 
horizon\footnote{For charged black holes, there exists inner horizons also.}, where $g_{tt}$ has a first order zero and $g_{rr}$ has a first order pole. We also
 assume that all the other metric components are finite as well as non vanishing at the horizon. The boundary of the space time is at $r \to \infty$. 

Since our aim is to compute the electrical conductivity using Kubo formula, it is sufficient to consider perturbations in the tensor (metric)  and the 
vector (gauge fields) modes around the black hole solution and keep other fields such as scalars unperturbed. We consider the perturbations of the form
\begin{equation}
g_{\mu\nu}= {\bf g}^{(0)}_{\mu\nu} + h_{\mu\nu}~,
\quad\quad
A_\mu = {\bf A}^{(0)}_\mu + {\cal A}_\mu~,
\end{equation}
where ${\bf g}^{(0)}_{\mu\nu}$ and ${\bf A}^{(0)}_\mu $ are background metric and gauge fields.

In order to determine electrical conductivity it is enough to consider perturbations in $(tx^{1})$ and
$(x^{1}x^{1})$ components of the metric tensor and $x^{1}$ component of the gauge fields. Moreover one can
choose the perturbations to depend on radial coordinate $r$, time $t$ and one of the spatial coordinate say $x^{2}$. %\footnote{We shall work with perturbations where rotational symmetry and the translational symmetry along $t,x^{i}$ is maintained.} $x^{2}$. 
A convenient ansatz with the above restrictions in mind is
\begin{equation}
h_{tx^{1}} = {\bf g}^{(0)}_{x^{1}x^{1}}~T(r)~e^{-i\omega t + i q x^{2}},\quad
h_{x^{2}x^{1}} = {\bf g}^{(0)}_{xx}~Z(r)~e^{-i\omega t + i q x^{2}},\quad
{\cal A}_{x^{1}} =  \phi(r)~~e^{-i\omega t + i q x^{2}}.\quad
\end{equation}
Here $\omega$ and $q$ represent the frequency and momentum in $x^{2}$ direction respectively and we set perturbations in the other components to be equal to zero.
 Our next step is to find linearized equations which follow from the equations of motion. It turns out that at the level of linearized equation and at zero 
momentum limit metric perturbation $Z(r)$ decouples from the rest. One can further eliminate $T(r)$ reducing it to equation for perturbations in gauge fields only. 
After substitution one finds the equation for perturbed gauge field to be
\begin{equation}
\frac{d}{dr}(N(r) \frac{d}{dr}\phi(r))-\omega^2 N(r)~ g_{rr} g^{tt} \phi(r)+  M(r) \phi(r)=0,
\label{eqnmotion}
\end{equation}
with
\begin{equation}
N(r)=\sqrt{-g}\frac{1}{g_{\rm{eff}}^2}g^{xx}g^{rr},\label{N}
\end{equation}
and
\begin{equation}
 M(r)= \Big(\frac{1}{g_{\rm{eff}}^2}\Big)^2 \sqrt{-g} g^{xx}g^{rr}g^{tt}F_{rt}F_{rt}.
\end{equation}
We can rewrite $M(r)$, in a better way as
 \begin{equation}
 M(r) = (2\kappa^{2})^{2} \rho^{2} \frac{g_{rr} g_{tt}}{\sqrt{-g} g_{xx}}.\label{eqnmotion0}
\end{equation}
where,
\begin{equation}
\rho = \frac{1}{2 \kappa^{2}g_{\rm{eff}}^2} \sqrt{- g}g^{rr} g^{tt} F_{r t} .\label{rho}
\end{equation} 
Let us note that the Maxwell equations can be written as,
\begin{equation}
\partial_\mu \Big(\frac{1}{g_{\rm{eff}}^2}\sqrt{- g}F^{\nu \mu}\Big) = 0,
\end{equation} 
and we choose the gauge where only  $A_{t}(r)$ component of the background gauge field is   is non zero (we work with electrically charged black hole).

For evaluating the conductivity in the low frequency limit and for non-extremal backgrounds, we only need to solve equations up to zeroth order in $\omega$. To that order one finds,
 \begin{equation}
\frac{d}{dr}(N(r)\frac{d}{dr}\phi(r))+ M(r)\phi(r)=0.\label{eqnmotion1}
 \end{equation}
The expression for electrical conductivity is given by (see \cite{Jain:2010ip, Jain:2009pw, Myers:2009ij} and Appendix A for details ),
\begin{eqnarray}
\sigma &=& \frac {1}{2\kappa^{2}} \Bigg(\sqrt{\frac{g_{rr}}{g_{tt}}}  N(r)\Bigg)_{r=r_{h}} \Bigg(\frac{\phi(r_{h})}{\phi(r\rightarrow \infty)}\Bigg)^{2} \nonumber\\
 &=& \frac {1}{2 \kappa^{2}}\Bigg( \frac{1}{g_{\rm{eff}}^2}~ g_{xx}^\frac{d-3}{2} \Bigg)_{r=r_{h}}\Bigg(\frac{\phi(r_{h})}{\phi(r\rightarrow \infty)}\Bigg)^{2}\nonumber\\
&=& \sigma_{H}~~\Bigg(\frac{\phi(r_{h})}{\phi(r\rightarrow \infty)}\Bigg)^{2} ,\label{sincleconduc}
\end{eqnarray} 
where $\sigma_{H} $ is the conductivity evaluated at the horizon and its expression is given by,
\begin{equation}
\sigma_{H} = \frac {1}{2 \kappa^{2}g_{\rm{eff}}^2}~ g_{xx}^\frac{d-3}{2}\Big{|}_{r=r_{h}}.\end{equation}

It was proposed in \cite{Jain:2010ip}, that
\begin{eqnarray}
\sigma &=& \sigma_{H}~~\Bigg(\frac{\phi(r_h)}{\phi(r\rightarrow \infty)}\Bigg)^{2}\nonumber\\
&=&\sigma_{H} \Bigg(\frac{s T}{\epsilon + P}\Bigg)^{2} ,\label{proposal}
\end{eqnarray} and
\begin{equation}
 \frac{\phi(r)}{\phi(r_{h})} = 1+\frac{\rho}{s T} (A_{t}(r) - A_{t}(r_{h})),\label{sol}
\end{equation} where  Eq (\ref{sol}) at the boundary reduces to 
\begin{eqnarray}
 \frac{\phi(r \rightarrow \infty)}{\phi(r_{h})} &=& 1+\frac{\rho}{s T} \mu \nonumber\\
&=& \frac{\epsilon+P}{s T}.\label{sol1}
\end{eqnarray}
So in order to show Eq (\ref{proposal}) we need to prove Eq (\ref{sol}). 
In the next section we show that Eq (\ref{sol}) indeed, is the solution to Eq (\ref{eqnmotion1}). 

\section{Proof} 

In this section we want to prove Eq (\ref{proposal}) or in other words, given a gravity background we want to understand under what 
conditions, the dual gauge theory will show behavior as in  Eq (\ref{proposal}).
The way we shall proceed is, first we shall assume that the solution to Eq (\ref{eqnmotion1}) is given by Eq (\ref{sol}).
Then we shall use Einstein equation to find out the constraint that our assumption leads to and we obtain these constraints can be expressed in a compact form in terms of the stress energy momentum tensor of the matter content of the system.

We start by plugging  Eq (\ref{sol}) in Eq (\ref{eqnmotion1}). This gives,
\begin{equation}
\frac{d}{dr}\left(\sqrt{-g}\frac{1}{g_{\rm{eff}}^2}g^{xx}g^{rr}\frac{\rho}{s T}\frac{d}{dr}A_{t}(r)\right) + (2\kappa^{2})^{2} \rho^{2} \frac{g_{rr} g_{tt}}{\sqrt{-g} g_{xx}} \left(1+\frac{\rho}{s T} \big(A_{t}(r) - A_{t}(r_{h})\big)\right)= 0.\nonumber
\end{equation} Using $F_{rt}=\frac{d}{dr}A_{t}$ and definition of charge density as in Eq (\ref{rho}) we obtain
\begin{eqnarray}
&&  2 \kappa^{2} \frac{\rho^{2}}{s T}\frac{d}{dr}(g^{xx} g_{tt}) +(2\kappa^{2})^{2} \rho^{2} \frac{g_{rr} g_{tt}}{\sqrt{-g} g_{xx}} \left(1+\frac{\rho}{s T} \big(A_{t}(r) - A_{t}(r_{h})\big)\right)= 0, \nonumber\\
\rm{or,}&&  \frac{1}{2 \kappa^{2}}\frac{\sqrt{-g} g_{xx}}{g_{rr} g_{tt}}\frac{d}{dr}(g^{xx} g_{tt}) = - s T \left(1 + \frac{\rho}{s T} \big(A_{t}(r) - A_{t}(r_{h})\big)\right).\label{tobep}
\end{eqnarray} 
Evaluating Eq (\ref{tobep}) at $r = r_{h}$,we get
\begin{equation}
 \frac{1}{2 \kappa^{2}}\frac{\sqrt{-g} g_{xx}}{g_{rr} g_{tt}}\frac{d}{dr}(g^{xx} g_{tt})\Bigg{|}_{r_{h}} = - s T .\label{tobeph}
\end{equation} Subtracting Eq (\ref{tobep}) from Eq (\ref{tobeph}) we get
\begin{eqnarray}
&& \frac{\sqrt{-g} g_{xx}}{g_{rr} g_{tt}}\frac{d}{dr}(g^{xx} g_{tt})\Bigg{|}_{r_{h}}^{r} =- 2 \kappa^{2} \rho \big(A_{t}(r) - A_{t}(r_{h})\big)\nonumber\\
&\Rightarrow &\Bigg[\frac{g_{xx}^{\frac{d+1}{2}}}{g_{tt}^{\frac{1}{2}} g_{rr}^{\frac{1}{2}}}\frac{d}{dr}(g^{xx} g_{tt})\Bigg{]}_{r_{h}}^{r} = -2 \kappa^{2} \rho A_{t}\Bigg{|}^{r}_{r_{h}}.\label{tobep1}
\end{eqnarray}
Now we use Einstein equations to find out conditions under which Eq (\ref{tobep1}) is valid. 
Let us consider the background of the form given in Eq (\ref{met1}). The Einstein equation is given by
\begin{eqnarray}
 R_{\mu \nu}-\frac{1}{2}g_{\mu \nu} R &=& T^{E.M.}_{\mu\nu}+T^{Matter}_{\mu\nu}\nonumber\\
&=& \frac{1}{2g_{\rm{eff}}^2} \left(F_{\mu \lambda} F_{\nu}^{\lambda} - \frac{1}{4} g_{\mu \nu} F_{\rho \sigma} F^{\rho \sigma}\right) + T^{Matter}_{\mu \nu},\label{ein}
\end{eqnarray} where $T^{Matter}_{\mu \nu}(r)$, will include all the other stuffs which may come from scalar fields, cosmological constant or any other 
fields present in the theory. 
Since only $A_{t}(r)$ is non-zero, we have $F_{rt}\neq 0.$  
Using Eq (\ref{ein}), we can write 
\begin{equation}
 R_{t}^{t}-\frac{1}{2}g_{t}^{t} R = \frac{1}{2g_{\rm{eff}}^2} \left(F_{tr} F^{tr} - \frac{1}{4} g_{t}^{t} F_{\rho \sigma} F^{\rho \sigma}\right) + T^{t,~Matter}_{t} \label{rtt},
\end{equation}
\begin{equation}
 R_{x}^{x}-\frac{1}{2}g_{x}^{x} R = -\frac{1}{2g_{\rm{eff}}^2}  \frac{1}{4} g_{x}^{x} F_{\rho \sigma} F^{\rho \sigma} +  T^{x,~Matter}_{x} \label{rxx}.
\end{equation}
After subtracting Eq (\ref{rtt}) from Eq (\ref{rxx}),  we get 
\begin{equation}
 \sqrt{-g} R^{t}_{t}-\sqrt{-g} R^{x}_{x} = \frac{1}{2g_{\rm{eff}}^2} \sqrt{-g} F^{rt} F_{rt} + \sqrt{-g}( T^{t,~Matter}_{t}(r)-T^{x,~Matter}_{x}(r))\label{rxt}.
\end{equation}
For the metric of the form in Eq (\ref{met1}), following relations hold
\begin{equation}
 \sqrt{-g} R^{t}_{t} = - \frac{d}{dr}\left(\frac{g_{xx}^{\frac{d-1}{2}}\frac{d}{dr}g_{tt}}{2 g_{rr}^{\frac{1}{2}}g_{tt}^{\frac{1}{2}}}\right),
\end{equation}
\begin{equation}
\sqrt{-g} R^{x}_{x} = - \frac{d}{dr}\left(\frac{g_{xx}^{\frac{d-3}{2}}g_{tt}^{\frac{1}{2}}}{2 g_{rr}^{\frac{1}{2}}}\frac{d}{dr}g_{xx}\right),
\end{equation} which, after substituting in Eq (\ref{rxt}), we get,
\begin{eqnarray}
   - \frac{d}{dr}\left(\frac{g_{xx}^{\frac{d-1}{2}}}{2 g_{rr}^{\frac{1}{2}}g_{tt}^{\frac{1}{2}}}\frac{d}{dr}g_{tt}\right) +\frac{d}{dr}\left(\frac{g_{x}^{\frac{d-3}{2}}g_{tt}^{\frac{1}{2}}}{2 g_{rr}^{\frac{1}{2}}}\frac{d}{dr}g_{xx}\right) &=& \frac{1}{2 g_{\rm{eff}}^2} \sqrt{-g} F^{rt} F_{rt}\nonumber\\
&+&\sqrt{-g}( T^{t,Matter}_{t}-T^{x,Matter}_{x}).
\end{eqnarray}
Upon further simplification, this reduces to
\begin{eqnarray}
-\frac{d}{dr}\left(\frac{g_{xx}^{\frac{d+1}{2}}}{g_{tt}^{\frac{1}{2}} g_{rr}^{\frac{1}{2}}}\frac{d}{dr}(g^{xx} g_{tt})\right) = 2 \kappa^{2} \rho \frac{d}{dr} A_{t}+ 2 \sqrt{-g}( T^{t,~Matter}_{t}(r)-T^{x,~Matter}_{x}(r)).
\end{eqnarray} 
Integrating above equation we get
\begin{equation}
 \left(\frac{g_{xx}^{\frac{d+1}{2}}}{g_{tt}^{\frac{1}{2}} g_{rr}^{\frac{1}{2}}}\frac{d}{dr}(g^{xx} g_{tt})\right)\Bigg{|}_{r_{h}}^{r} =-  2 \kappa^{2} \rho A_{t}\Bigg{|}^{r}_{r_{h}}+2\int_{r_{h}}^{r} dr \sqrt{-g}( T^{t,~Matter}_{t}(r)-T^{x,~Matter}_{x}(r))\label{pc}. 
\end{equation} 
Thus, if we impose the condition that 
\begin{equation}
T^{t,~Matter}_{t}(r) = T^{x,~Matter}_{x}(r), \label{condition}
\end{equation}
 then we get
\begin{equation}
  \left(\frac{g_{xx}^{\frac{d+1}{2}}}{g_{tt}^{\frac{1}{2}} g_{rr}^{\frac{1}{2}}}\frac{d}{dr}(g^{xx} g_{tt})\right)\Bigg{|}_{r_{h}}^{r} = - 2 \kappa^{2} \rho A_{t}\Bigg{|}^{r}_{r_{h}}\label{epst},
\end{equation}
 which\footnote{For the backgrounds which satisfies Eq (\ref{condition}), it is interesting to note that,  if we set $r\rightarrow \infty$, and use first law of thermodynamics as well as the fact that $s T_{H} = \frac{1}{2\kappa^{2}}\left(\frac{g_{xx}^{\frac{d+1}{2}}}{g_{tt}^{\frac{1}{2}} g_{rr}^{\frac{1}{2}}}\frac{d}{dr}(g^{xx} g_{tt})\right)\Bigg{|}_{r_{h}},$  we have $\epsilon + P = \frac{1}{2\kappa^{2}}\left(\frac{g_{xx}^{\frac{d+1}{2}}}{g_{tt}^{\frac{1}{2}} g_{rr}^{\frac{1}{2}}}\frac{d}{dr}(g^{xx} g_{tt})\right)\Bigg{|}_{r\rightarrow \infty}$ from Eq (\ref{tobep1}). Let us note that we should add the Gibbons-Hawking term and counter terms (see \cite{Batrachenko:2004fd}) in order to get finite values.} is same as Eq (\ref{tobep1}). Hence, what we have shown is, if the gravity background satisfies Eq (\ref{condition}), then the 
dual gauge theory will satisfy  Eq (\ref{sincleconduc}). 
We suspect that whenever the boundary theory is in the Minkowski space, the condition imposed by Eq (\ref{condition}) on the stress-energy tensor (barring the electromagnetic part) will hold true. This was also observed in \cite{Buchel:2003tz, Buchel:2004qq} in the context of proving the universality of shear viscosity.  In the following section, we elaborate upon the above condition considering several examples.

\section{Examples}

In all of our examples in this section we will take the metric, gauge fields and other form fields as the functions of coordinate $r$ only. It was observed in \cite{Buchel:2003tz, Buchel:2004qq} that if the scalar and other form fields are functions of the coordinate $r$ only and if the boundary theory lives on the Minskowski space, then  $T_{\mu\nu}^{~~\rm{Matter}}\sim g_{\mu\nu}(\cdots)$, which in turn implies the condition given by Eq (\ref{condition}). In what follows, in this section, we first discuss the boundary theories which live on Minkowski space-time where we will find explicitly that the Eq (\ref{condition}) holds good. Next, we discuss one example where the boundary theory does not live on the Minkowski space-time, namely the asymptotically Lifshitz like space-time, where the condition does not hold.  

% Moreover, the result 
% in Eq (\ref{sincleconduc}) is valid for the backgrounds which satisfies Eq (\ref{condition}). 
% %In general, if the 
% different scalar fields in gravity side are $\psi_{i}(r)$, then the form of
% $T^{Matter}_{\mu\nu}$ is given by,
% \begin{equation}
% T^{Matter}_{\mu\nu} = \sum_{i,j} G_{ij}(\psi)\d_{\mu}\psi_{i}(r) \d_{\nu}\psi_{j}(r)+ g_{\mu\nu}(r) V(\psi).
% \end{equation} It can be easily checked that $T^{t,Matter}_{t} = T^{x,~Matter}_{x}$, but $T^{t,~Matter}_{t} \neq T^{r,~Matter}_{r}$.  

\begin{itemize}
 \item \textbf{Boundary theories living on Minskowski space-time}
\begin{itemize}
 \item \textbf{Conformal boundary theories:} Let us note that  Reissner Nordstr\"{o}m and R-charged black holes in  various dimensions in asymptotically AdS space (as already checked in \cite{Jain:2010ip}) as well as any other background which satisfies Eq (\ref{condition}),
  should satisfy Eq (\ref{sincleconduc}).

\item\textbf{Non-conformal boundary theory:} Non-conformal theories such as gauge theory dual to charged $Dp$ brane satisfies
Eq (\ref{sincleconduc}). We shall check this explicitly in the next section.
\end{itemize}
% \item \textbf{Boundary theories which does not live on Minkowski space-time}
% \begin{itemize}
\item\textbf{Boundary theory dual to charged Lifshitz like black hole:} For this case it was computed in \cite{Jain:2010ip}
 that 
\begin{equation}
 \sigma_{B} \neq \sigma_{H} \Big(\frac{s T}{\epsilon + P}\Big)^{2}.  \label{ineq}
\end{equation} Now the above result can be understood easily.
Let us consider the following action in $(d+2)$-dimensional space time (see for details in \cite{Taylor:2008tg,Pang:2009pd})
\begin{equation}
S=\frac{1}{16\pi G_{d+2}}\int
d^{d+2}x\sqrt{-g}(R-2\Lambda-\frac{1}{4}F^{2}-\frac{1}{2}m^{2}A^{2}-\frac{1}{4}F_{1}^{2}).
\end{equation}
The corresponding equations of motion are given as follows,
\begin{eqnarray}
\partial_{\mu}(\sqrt{-g}F^{\mu\nu})&=&m^{2}\sqrt{-g}A^{\nu},~~~
\partial_{\mu}(\sqrt{-g}F_{1}^{\mu\nu})=0,\nonumber\\
R_{\mu\nu}&=&\frac{2}{d}\Lambda
g_{\mu\nu}+\frac{1}{2}F_{\mu\lambda}{F_{\nu}}^{\lambda}
+ \frac{1}{2} F_{1, \mu\lambda}F_{1,\nu}^{\lambda}+\frac{1}{2}m^{2}A_{\mu}A_{\nu}\nonumber\\
&
&-\frac{1}{4d}F^{2}g_{\mu\nu}-\frac{1}{4d}F_{1}^{2} g_{\mu\nu}\label{rmn}.
\end{eqnarray}
From the above equation we can find the energy momentum tensor. Let us write it in the form $T^{total}_{\mu\nu} = T^{E.M.}_{\mu\nu}+T^{Matter}_{\mu\nu},$ where $T^{E.M.}_{\mu\nu}$ contains contribution from gauge field $F_{1, \mu\nu}$ whereas other fields contributes to $T^{Matter}_{\mu\nu}.$ Let us note that the massive gauge field $A_{\mu}$; was introduced to get the Lifshitz like scaling.
% The perturbation that we will take is in the field $F_{1, \mu \nu}$. So that comparing Eq (\ref{rmn}) with Eq (\ref{ein}), we conclude
% \begin{equation}
%  T^{Matter}_{\mu \nu} = \frac{2}{d}\Lambda
% g_{\mu\nu}+\frac{1}{2}F_{\mu\lambda}{F_{\nu}}^{\lambda}
% +\frac{1}{2}m^{2}A_{\mu}A_{\nu}-\frac{1}{4d}F^{2}g_{\mu\nu}.
% \end{equation}
If we take only non-vanishing components of gauge field to be $A_{t}$, then it is easy to see that
\begin{eqnarray}
 T^{t,Matter}_{t} - T^{x,Matter}_{x} & = &\frac{1}{2}F_{t r}F^{t r } +\frac{1}{2}m^{2}A_{t}A^{t}\nonumber\\
& \neq & 0 ,
\label{lifpro}
\end{eqnarray} where $F_{rt} = \frac{d}{dr} A_{t}$ and also note that $g^{t}_{t}=g^{x}_{x} =1$.
This provides us with the explanation of Eq (\ref{ineq}). 
\end{itemize}

\section{Away from conformality:}  We shall motivate the purpose of this section by giving example of single charged $D1$ brane case recently considered in \cite{David:2010qc}. Then we shall check explicitly the validity of Eq (\ref{sincleconduc}) for non-conformal
gauge theories dual to general charged $Dp$ brane as well as  give general results for multiple charged $Dp$ brane.

 \begin{itemize}
  \item \textbf{Electrical conductivity for charged $D1$ brane:}
Let us consider the following action
 \begin{eqnarray}
 I & =& \frac{1}{16\pi G_3} \int d^3 x \sqrt{-g} \Big[ R(g) - \frac{8}{9} \partial_\mu \phi \partial^\mu \phi 
-\frac{1}{4} \Psi^2 e^{-\frac{4}{3}\phi} F_{\mu\nu} F^{\mu\nu} \nonumber \\
&-&  \frac{1}{2\Psi^2} \partial_{\mu} \Psi \partial^\mu \Psi + \frac{2}{3\Psi}\partial_{\mu}{\phi}\partial^{\mu}\Psi  +\frac{12}{L^2} e^{\frac{4}{3}\phi} (1+ {\Psi}^{-1})\Big]  \label{truncact}.
\end{eqnarray} 
% where
% \begin{equation}
% \frac{1}{G_{3}} = \frac{ 2\pi^4 L^7} { 3! G_{10}}, \qquad
% G_{10} = 2^3 \pi^6 g_s^2 (\alpha')^4.
% \end{equation}
Here again one can easily check that Eq (\ref{condition}) is satisfied, so here our general formula Eq (\ref{sincleconduc}) must hold.
In the following we shall check this explicitly. 

The metric, gauge field and scalar fields are given by 
 \begin{eqnarray}
 \label{finaltruncsol}
ds^2 &=& \left( -c_T^2 dt^2 + c_X^2 dz^2  + c_R^2 dr^2\right) ,\\ \nonumber
c_T^2 &=& \left( \frac{r}{L} \right)^8 K, 
\qquad
c_X^2 = \left( \frac{r}{L} \right)^8 H,\qquad c_R^2 = \frac{H}{K} \left( \frac{r}{L}\right)^2,  \\ \nonumber
A_t &=& - \frac{r_0^3 l}{ L^2  ( r^2 + {l^2}) }, \qquad \phi = -3 \log \left( \frac{r}{L} \right) ,\qquad
\Psi = 1 + \frac{l^2}{r^2}.
\end{eqnarray}
Here $H$ and $K$ are defined as
\begin{equation}
\label{defhk}
H = 1 + \frac{l^2}{r^2}, \qquad K = 1 + \frac{l^2}{r^2} - \frac{r_0^6}{r^6}.
\end{equation}
Different thermodynamic quantities are given by,
\begin{equation}
\label{therm1}
 T = \frac{1}{2\pi L^3} \frac{r_H^5}{r_0^3} ( 3+ 2k), \qquad s = \frac{1}{ 4 G_3} \frac{ r_0^3 r_H}{L^4},
\end{equation}
where $k$  is given by 
\begin{equation}
\label{defkl}
 k = \frac{l^2}{r_H^2},
\end{equation}
 and $r_H$ is the radius of the horizon which is given by the largest root of the  equation
\begin{equation}\label{r0rH}
 r_H^6 + r_H^4 l^2 - r_0^6 =0.
\end{equation}
The energy density ($\epsilon$) and the pressure ($p$) is given by
\begin{equation}
\label{therm2}
 \epsilon = \frac{1}{4\pi G_3} \frac{ r_0^6}{L^7}, \qquad p =  \frac{1}{8\pi G_3} 
\frac{ r_0^6}{L^7}=  \frac{\epsilon}{2}.\end{equation}The charge density $\rho$  and its conjugate the  chemical potential $\mu$  are given by
\begin{equation}
\label{therm3}
 \rho = \frac{1}{8\pi G_3} \frac{ r_0^3 l}{ L^5} ,\qquad \mu = 
 A_{t} (r) |_{r\rightarrow\infty}- A_t(r)|_{r_H}=  \frac{ l r_H^4}{ L^2 r_0^3}.
\end{equation}
So conductivity should be,
\begin{eqnarray}
 \sigma &=&  \frac{1}{16\pi G_3} \frac{1}{g_{\rm{eff}}^2} g_{xx}^{-\frac{1}{2}}\Big{|}_{r=r_{h}} \Big(\frac{s T}{\epsilon + P}\Big)^{2}\nonumber\\
 &=& \frac{1}{16\pi G_3}\Psi^2 e^{-\frac{4}{3}\phi} g_{xx}^{-\frac{1}{2}}\Big{|}_{r=r_{h}} \Big(\frac{s T}{\epsilon + P}\Big)^{2}\nonumber\\
&=&\f{1}{16\pi G_{3}}\frac{(2k+3)^2}{9\sqrt{1+k}},
\end{eqnarray} which is same as the one computed in \cite{David:2010qc}. In that paper authors also computed electrical conductivity for four equal charge case. 
The results follow from Eq (\ref{sincleconduc}) in a straight forward manner. 

\item \textbf{Thermal conductivity :} In \cite{David:2010qc}, thermal conductivity to bulk viscosity ($\zeta$) ratio for both single charge and equal four charge case was
 computed to be
\begin{equation}
 \frac{\kappa_T}{\zeta T} \mu^{2} = 4\pi^{2} L^{2}.\label{D1kt}
\end{equation} 
\item Recently it was noted in \cite{Jain:2009bi,Jain:2010ip}, that this particular ratio remains unchanged
 even if we take multiple chemical potentials  or set chemical potential to zero. In the following we check  this again.
\end{itemize}

\subsection{Uncharged $Dp$ brane:}
Let us consider the following background
\begin{equation}
 ds^{2} = -g^{\frac{n+1}{d-2}} r^{\frac{n+1}{d-2}}f(r) dt^{2} + g^{\frac{n+1}{d-2}} r^{\frac{n+1}{d-2}} \sum_{i =1}^{p} dx_{i}^{2} + g^{1-n+\frac{n+1}{d-2}} r^{1-n + \frac{n+1}{d-2}}\frac{1}{f(r)} dr^{2}, \label{unchargedmetric}
\end{equation} where $f(r) = 1-\frac{2 m}{r^{n-1}}$, $d = p+2$, $n = 10-d$ and $g$ is a constant ($=\frac{1}{L}$, in the notation used in \cite{David:2010qc} ).
 Note that this background is exactly same as the one considered in \cite{Kovtun:2003wp}. We have written it differently.
As it was argued in \cite{Kovtun:2003wp}, in order to determine electrical conductivity one 
should consider the following Maxwell equation,
\begin{equation}
 S_{Maxwell} = -\frac{1}{16 \pi G}\int dx^{p+2}\sqrt{-g_{p+2}}\frac{1}{4~g_{eff}^{2}}F_{\mu \nu} F^{\mu \nu},
\end{equation} where
\begin{equation} g_{eff}^{2} = (g r)^{-\frac{a^{2} (D-2)}{2(d-2)}},~~ a^{2} = 4-2\frac{(n-1)(D-n-1)}{D-2},\end{equation} and 
$D$ is the higher dimensional space from where we are reducing the metric down to $d$ dimensions ( for our case $D= 10$).
Various thermodynamic quantities are given by,
\begin{equation}
 T = \frac{n-1}{4\pi r_{h}}(g r_{h})^{\frac{n-1}{2}},~~  s = \frac{1}{4 G} g_{xx}^{\frac{p}{2}}\Big{|}_{r_H},
\end{equation} and $\epsilon + P = s T$. The electrical conductivity is given by 
\begin{eqnarray}
 \sigma &=& \frac{1}{16 \pi G} \frac{1}{g_{eff}^{2}} g_{xx}^{\frac{p-2}{2}}\Big{|}_{r_H}\nonumber\\
&=& \frac{1}{16 \pi G} (g r_{h})^{\frac{7-n}{2}}.
\end{eqnarray}
It is easy to see that,
\begin{eqnarray}
 D_{R} &=& \frac{\sigma}{\chi}\nonumber\\ &=& \frac{7-p}{8 \pi T},
\end{eqnarray}  as was shown in \cite{Kovtun:2003wp}, where 
\begin{eqnarray}
 \chi &=& \frac{\rho}{\mu}\nonumber\\
&=& \frac{1}{8\pi G} g^{3} r_{h}^{2}\label{chichi}.
\end{eqnarray}
 Let us note that, though $\rho$ and $\mu$  go to zero separately for uncharged $Dp$ brane, $\chi$ in Eq (\ref{chichi}), remains non-zero.
Now  using expression for thermal conductivity,  $\kappa_T = \frac{(\epsilon + p)^{2} \sigma}{\rho^{2} T}, $ we get
 \begin{eqnarray}
 \frac{\kappa_T}{\eta T} \mu^{2} &=& 4 \pi \left(\frac{\sigma}{\chi}\right)^{2} \frac{s}{\sigma}\nonumber\\
&=& \frac{4 \pi^{2}}{g^{2}}.\label{unchargedkt}
\end{eqnarray}
Note that, from Eq (\ref{unchargedkt}), we see that thermal conductivity to viscosity ratio is same for any uncharged 
$Dp$ brane. Also note, to match with charged $D1$ brane, replace $\eta$
 by bulk viscosity and
 $g =\frac{1}{L}$.

Our next aim is to see whether for charged non-conformal theories dual to charged $Dp$ brane, thermal conductivity to
viscosity ratio remains $\frac{4 \pi^{2}}{g^{2}}$. 
\subsection{Charged $Dp$ brane}
Let us consider the background obtained from Kaluza-Klein spherical reduction of single charged rotating black $Dp$
 brane to $d$ dimension (see for details \cite{Cvetic:2000zu,Harmark:1999xt,Cvetic:2000dm}).
\begin{equation}
 ds^{2} = -g^{\frac{n+1}{d-2}} r^{\frac{n+1}{d-2}} h^{-\frac{d-3}{d-2}} f(r) dt^{2} + g^{\frac{n+1}{d-2}} r^{\frac{n+1}{d-2}} h^{\frac{1}{d-2}} \sum_{i =1}^{p} dx_{i}^{2} + g^{1-n+\frac{n+1}{d-2}} r^{1-n + \frac{n+1}{d-2}} h^{\frac{1}{d-2}}\frac{1}{f(r)} dr^{2}, \label{chargedmetric}
\end{equation}
% \begin{equation}
%  g_{tt} = -g^{\frac{n+1}{d-2}} r^{\frac{n+1}{d-2}} h^{-\frac{d-3}{d-2}} f(r),~~ g_{xx} = g^{\frac{n+1}{d-2}} r^{\frac{n+1}{d-2}} h^{\frac{1}{d-2}},~~~~ g_{rr} = g^{1-n+\frac{n+1}{d-2}} r^{1-n + \frac{n+1}{d-2}} h^{\frac{1}{d-2}}\frac{1}{f(r)},
% \end{equation}
 where 
\begin{equation}
 f(r) = h-\frac{2 m}{r^{n-1}}, ~~~ h = \prod_{i =1}^{b}(1 + H_{i}), ~~~~~H_{i} = 1+ \frac{l_{i}^{2}}{r^{2}},
\end{equation} where $b$ is the number of independent gauge fields (which is same as number of independent spins
 that a higher dimensional $Dp$ brane can have before compactification).
The action is of the form
\begin{equation}
 S = \frac{1}{16 \pi G}\int d^{p+2}x \sqrt{-g}\Big[R -\frac{1}{4}\sum_{i=1}^{b}\frac{1}{X_{i}^{2}} F_{\mu\nu}^{i}F^{i~\mu\nu}+ all~the~other~ terms..\Big],
\end{equation} where 
\begin{equation}
 X_{i} = g^{-\frac{a^{2}(D-2)}{4(d-2)}} r^{-\frac{a^{2}(D-2)}{4(d-2)}} h^{\frac{d-3}{2(d-2)}} \frac{1}{H_{i}},
\end{equation} and
\begin{equation}
 A_{t}^{i} =-\sqrt{2 m} g^{\frac{n-3}{2}}\frac{1-\frac{1}{H_{i}}}{l_{i}}. 
\end{equation}
In the following we define all the required thermodynamic quantities. The expression for charge density is,
\begin{equation}\rho_{i}= \frac{1}{8\pi G}\sqrt{2 m} g^{\frac{n+3}{2}} l_{i},\end{equation} the chemical potentials are given by
\begin{equation}
 \mu_{i}=\sqrt{2 m} g^{\frac{n-3}{2}}\frac{l_{i}}{r_{h}^{2}H_{i}(r_{h})}.\end{equation} The Hawking temperature is given by
\begin{equation} 
T=\frac{\sqrt{m}}{\sqrt{2}\pi r_{h}}g^{\frac{n-1}{2}}(\frac{n-1}{2}-\frac{1}{r_{h}^{2}}\sum_{j=1}^{b}\frac{l_{i}^{2}}{H_{i}(r_{h})}).\end{equation}
The expression for entropy and other required quantities are
\begin{equation}
s = \frac{1}{4G}g^{\frac{n+1}{2}}r_{h}\sqrt{2 m} ,~~\epsilon+P = \frac{(n-1) m}{8\pi G}g^{n}.
\end{equation} 
The equation\footnote{Unless explicitly mentioned, there is no sum over repeated indices $i,j$.} that we have to solve in order to find out conductivity is given by
\begin{equation}
\frac{d}{dr}(N_i\frac{d}{dr}\phi_i(r)) + \sum\limits_{j=1}^m  M_{ij}\phi_j(r)=0.
\label{eqnmotionmulti}
\end{equation} where
\begin{equation}
N_i=\sqrt{-g}\frac{1}{X_{i}^{2}}g^{xx}g^{rr},\label{Ni}
\end{equation}
and
\begin{equation}
 M_{ij}=F_{rt}^i \sqrt{-g}\frac{1}{X_{i}^{2}}g^{xx}g^{rr}g^{tt}\frac{1}{X_{j}^{2}}F_{rt}^j.
\end{equation}
Plugging the background values we can show
\begin{equation}
N_{i} = g^{3}r^{3} f(r)H_{i}^{2}\frac{1}{h},~~~M_{ij} = -8 m~ l_{i}~l_{j}g^{3}r^{-n} \frac{1}{h}.
\end{equation}
\begin{itemize}
 \item \textbf{Single charge case:} Here we have 
\begin{equation}
 \sigma = \frac{1}{16\pi G}\frac{1}{X^{2}}g_{xx}^{\frac{p-2}{2}}|_{r_H} \Big(\frac{s T}{\epsilon + P}\Big)^{2}.
\end{equation} Next using the fact that,
\begin{equation}
 \frac{\rho}{\mu} = \frac{1}{8\pi G} g^{3} r_{h}^{2} H(r_{h}),
\end{equation} we get
\begin{eqnarray}
\frac{K_{T}\mu^{2} }{\eta T} &=& \frac{4\pi^{2}}{g^{2}}, 
\end{eqnarray} which is same as we get for uncharged case.
\item \textbf{Multicharge case:} For multicharge case, there is an analog of Eq (\ref{proposal}). As it was proposed in \cite{Jain:2010ip}, that for multicharge case 
\begin{equation}
 \rho_{i}\sigma^{-1}_{ij}\rho_{j} = \rho_{i}\sigma^{-1}_{H,ii}\rho_{i}\Bigg(\frac{\epsilon + P}{s T}\Bigg)^{2},\label{sigmat} 
\end{equation} where $\sigma^{-1}_{H,ii}$ is the inverse of electrical conductivity evaluated at the horizon and only depends on geometrical 
quantities evaluated at the horizon. The expression for electrical conductivity at the horizon is given by,
\begin{eqnarray}
 \sigma_{H,ii}  &=& \frac {1}{16\pi G} G_{ii}(r)~ g_{xx}^\frac{p-2}{2}\Big{|}_{r=r_{h}} \nonumber\\
&=& \frac {1}{16\pi G} \frac{1}{X_{i}^{2}}~ g_{xx}^\frac{p-2}{2}\Big{|}_{r=r_{h}}\nonumber\\
&=&  \frac{g^{\frac{7-n}{2}} r_{h}^{3}H_{i}^{2}(r_{h})}{16 \sqrt{2 m}~\pi G}\label{sighmat}.
\end{eqnarray} Using this result, it can be  easily shown that, \begin{equation}
 \frac{K_{T}\sum_{i=1}^{b}\mu_{i}^{2}}{\eta T} =\frac{4 \pi^{2}}{g^{2}}.
\end{equation}For $D1$ brane $\eta$ is replaced by $\frac{s}{4\pi}$ ( which is same as  bulk viscosity for single charge
case or equally charged $D1$ brane case as shown in \cite{David:2010qc}).
\item\textbf{ D1 brane with four unequal charges:}  In this case, the coupled set of equations for $i^{th}$ field are
given by
\begin{equation}
 \frac{d}{dr}(N_i\frac{d}{dr}\phi_i(r)) + \sum\limits_{j=1}^4  M_{ij}\phi_j(r)=0,
\end{equation} where index $i$, can take value from $1$ to $4$ (there is no sum over $i$ in the above) and 
\begin{equation}
N_{i} = g^{3}r^{3} f(r)H_{i}^{2}\frac{1}{h},~~~M_{ij} = -8 m~ l_{i}~l_{j}g^{3}r^{-7} \frac{1}{h},~~ h =  \prod_{i =1}^{4}(1 + H_{i}).
\end{equation} 
Demanding regularity (ingoing boundary condition) at the horizon and at the boundary $\phi_{i} = \phi^{0}_{i},$ we get the solution to $4$ coupled equation to be
\begin{equation}
\phi_{i} =\frac{ \phi^{0}_{i}+ \frac{l_{i}}{6 r^{2}}(6 l_{i} \phi^{0}_{i}-2\sum_{j=1}^{4} \phi^{0}_{j}l_{j})}{H^{2}_{i}}. 
\end{equation}
We can now compute the conductivity by using the expression discussed in reference \cite{Jain:2010ip}. The expression for diagonal part of electrical conductivity is given by
\begin{equation}\sigma_{ii}= \frac{9 r_{h}^{4}+12 r_{h}^{2}l_{i}^{2}+3 l_{i}^{4}+l_{i}^{2}\sum_{j=1}^{4}l_{j}^{2}}{144 \sqrt{2 m}~\pi G r_{h} },\end{equation}
whereas  off diagonal part of the conductivity is given by
\begin{equation}
\sigma_{ij}= -\frac{l_{i}l_{j}}{144 \sqrt{2 m}~\pi G r_{h}}(6 r_{h}^{2}+\sum_{k=1}^{4}l_{k}^{2}-3(l_{i}^{2}+l_{j}^{2})).
\end{equation}
 We can now explicitly check that, for multicharge case 
\begin{equation}
 \rho_{i}\sigma^{-1}_{ij}\rho_{j} = \rho_{i}\sigma^{-1}_{H,ii}\rho_{i}\Bigg(\frac{\epsilon + P}{s T}\Bigg)^{2},\label{sigmat1} 
\end{equation} where 
\begin{eqnarray}
 \sigma_{H,ii}  &=&  \frac{r_{h}^{3}H_{i}^{2}(r_{h})}{16 \sqrt{2 m}~\pi G}\label{sighmat1},
\end{eqnarray} is the electrical conductivity evaluated at the horizon and depends  only on the geometrical 
quantities evaluated at the horizon.
The thermal conductivity ($K_{T}$) can  be computed using Eq (\ref{sigmat}) and Eq (\ref{sighmat}), and 
\begin{equation}
 K_{T} = \left( \frac{\epsilon+P}{ T}\right)^2\frac{T}{\sum\limits_{i,j=1}^4\rho_{i}\sigma_{ij}^{-1}\rho_{j}}\label{thermalconductivity}.
\end{equation} Plugging all the expressions we get,
\begin{equation}
K_{T} =  \frac{\pi}{g^{2}}\frac{s T}{\sum_{i=1}^{4}\mu_{i}^{2}}.
\end{equation}

We now compute the required ratio
\begin{equation}
 \frac{K_{T}\sum_{i=1}^{4}\mu_{i}^{2}}{\frac{s}{4\pi} T} =\frac{4 \pi^{2}}{g^{2}}.
\end{equation}
\end{itemize}

\section{Conclusion}
In this paper we have shown that, for $\mu \neq 0$,  given that the form of 
 Maxwell part of the action is
\begin{equation} S = - \int d^{d+1}x~~\sqrt{-g}\frac{1}{4 g^{2}_{eff}}  F_{MN}F^{MN}\label{max2},\end{equation}
 the  electrical conductivity at the boundary is given by 
\begin{eqnarray}\sigma_{B} &=& \frac {1}{g^{2}_{eff}}  g_{xx}^\frac{d-3}{2}\Big{|}_{r=r_h} \frac{(sT)^{2}}{(\epsilon+ P)^{2}}\nonumber\\
&=& \sigma_{H}\frac{(sT)^{2}}{(\epsilon+ P)^{2}}\label{HBC},
\end{eqnarray} where $\sigma_{H} = \frac {1}{g^{2}_{eff}}  g_{xx}^\frac{d-3}{2}\Big{|}_{r=r_h}$,
 is the electrical conductivity evaluated radially at the horizon. Following \cite{Jain:2010ip}, we can argue that once the real part of 
the conductivity is known, the imaginary part of conductivity is automatically fixed. 
To summarize, in the presence of chemical potential 
the electrical conductivity  can be expressed as 
% \footnote{Let us note that, $\frac{g_{tt}}{g_{xx}}\Bigg|_{r \rightarrow \infty} = 1$, in most cases, so that\begin{equation}
%  \lambdabf = -\frac{1}{\omega}\frac{\rho^{2}}{\epsilon + P} + \frac{1}{ g^{2}_{eff}}~ g_{xx}^\frac{d-3}{2}\Big{|}_{r=r_h} \frac{(sT)^{2}}{(\epsilon+ P)^{2}}\nonumber.
% \end{equation}  }
\begin{equation}
 \lambdabf = -\frac{i}{\omega}\Bigg(\frac{g_{tt}}{g_{xx}}\Bigg)_{r \rightarrow \infty}\frac{\rho^{2}}{\epsilon + P} + \frac{1}{ g^{2}_{eff}}~ g_{xx}^\frac{d-3}{2}\Big{|}_{r=r_h} \frac{(sT)^{2}}{(\epsilon+ P)^{2}}.
\end{equation} Let us mention here that the imaginary part of the conductivity has a pole at $\omega\to 0$ limit because of the translational 
invariance of the system. If one uses the Krammers-Kronig relation
\begin{equation}
 \Im (\lambda(\omega)) =-\frac{1}{\pi}{\mathcal{P}}\int_{-\infty}^{\infty}\frac{\Re(\lambda(\omega'))}{\omega'-\omega}d\omega',
\end{equation}
then one can find that the real part of the conductivity contains a delta function iff the imaginary part has a pole. As we have found a pole in the imaginary part of the conductivity, it follows that real part has a delta function singularity at $\omega = 0.$ So, strictly speaking DC conductivity that we have computed is low frequency limit of AC conductivity or more precisely expression for conductivity is valid for  $w\to 0^{+},$ see \cite{Hartnoll:2009sz, David:2010qc} for a nice discussion.   

It is interesting to note that, following  
\cite{Jain:2010ip}, the cutoff dependent conductivity can be computed  which interpolates smoothly between the 
results at the horizon and at the boundary. At any cutoff $r_{c}$ the expression for electrical conductivity\footnote{ Let us note that, at any radius $r_{c}$, the local temperature and the chemical potential can be given by $T_{c} = \frac{T_{H}}{\sqrt{g_{tt}(r_{c})}}$ and $\mu_{c} = \frac{A_{t}(r_{c})-A_{t}(r_{h})}{\sqrt{g_{tt}(r_{c})}}$ respectively. Assuming first law of thermodynamics $\epsilon(r_{c})+ P(r_{c}) = sT_{c} + \rho \mu_{c}$ to hold at and radius and using Eq (\ref{sol}) we get\begin{equation}
 \frac{\phi(r_{c})}{\phi(r_{h})} =  \frac{sT}{\epsilon+ P}\Big{|}_{r=r_c} \nonumber,
\end{equation} and consequently Eq (\ref{rclam}).
 } 
can be written as 
\begin{equation}
 \lambdabf = -\frac{i}{\omega}\Bigg(\frac{g_{tt}}{g_{xx}}\Bigg)_{r_{c}} \Bigg(\frac{\rho^{2}}{\epsilon + P}\Bigg)_{r\rightarrow \infty} + \frac{1}{ g^{2}_{eff}}~ g_{xx}^\frac{d-3}{2}\Big{|}_{r=r_h} \frac{(sT)^{2}}{(\epsilon+ P)^{2}}\Big{|}_{r=r_c} \label{rclam},
\end{equation} where $r\rightarrow \infty$ is the boundary of the space time.
It is interesting to compare our results with the results obtained from the membrane paradigm arguments. We have seen, that
irrespective of the theory, the horizon conductivity is given by
\begin{equation}
 \sigma_{H} = \frac{1}{ g^{2}_{eff}}~ g_{xx}^\frac{d-3}{2}\Big{|}_{r=r_h}, 
\end{equation} whereas the universal conductivity of the membrane is given by
\begin{equation}
 \sigma_{membrane} = \frac{1}{ g^{2}_{eff}}~ \Big{|}_{r=r_h}.
\end{equation} So we conclude that the horizon conductivity is given by,
\begin{equation}
 \sigma_{H} = \sigma_{mem} g_{xx}^\frac{d-3}{2} \Big{|}_{r=r_h}.
\end{equation} We have also seen that for the background as of the form Eq (\ref{met1}), if Eq (\ref{condition}) is satisfied then
% \begin{equation}
%  R_{\mu \nu} = \frac{1}{2g_{\rm{eff}}^2} \left(F_{\mu \lambda} F_{\nu}^{\lambda} - \frac{1}{2 (d-1)} g_{\mu \nu} F_{\rho \sigma} F^{\rho \sigma}\right) + T^{Matter}_{\mu \nu}(r),\label{ein2}
% \end{equation} with 
% \begin{equation}
% T^{t,Matter}_{t} = T^{x,Matter}_{x},
% \end{equation}
 the boundary conductivity can be related to horizon conductivity using 
thermodynamic quantities. More precisely we can write,
\begin{eqnarray}
 \sigma_{B} &=& \sigma_{H} \frac{(sT)^{2}}{(\epsilon+ P)^{2}}\nonumber\\
&=& \sigma_{mem}~ g_{xx}^\frac{d-3}{2} \Big{|}_{r=r_h} \frac{(sT)^{2}}{(\epsilon+ P)^{2}}\label{HBc}.
\end{eqnarray} Since mass dimension of electrical conductivity is $d-3$, one can understand the factor $g_{xx}^{\frac{d-3}{2}}$ as the converter of the length scale of the boundary to the proper length at the horizon \cite{Iqbal:2008by,Myers:2009ij}. It would be very interesting to understand the meaning of extra factor $(\frac{s T}{\epsilon+ P})^{2}$ that appears in the formula due to presence of chemical potential. At this moment it is not quite clear to us how to interpret it directly  from the constraint Eq (\ref{condition}) which appears to be related to  Lorentz invariance of the vacuum of the field theory. Let us note that, at zero chemical potential
\begin{eqnarray}
 \sigma_{B} &=& \sigma_{H} \nonumber\\
&=& \sigma_{mem}~ g_{xx}^\frac{d-3}{2} \Big{|}_{r=r_h} ,
\end{eqnarray} as was shown in \cite{Iqbal:2008by}.

 In our result of electrical conductivity, $\sigma_{H}$ is given entirely in terms of gravity  theory. A natural question that arises,  whether  it is possible to give an intrinsic meaning to the  expression of conductivity  in terms of field theory quantities? This will put the formula for electrical conductivity in the same footing as celebrated universal result for $\frac{\eta}{s}$. Answer to this comes from the expression of thermal conductivity to viscosity ratio. As it was shown in \cite{Jain:2009pw}, electrical conductivity can be expressed in terms of the field theory quantities alone.

 We have  seen that, using universality of electrical conductivity, another universality of
thermal conductivity to shear viscosity ratio might be shown easily. We leave the proof of the universality of thermal conductivity to viscosity as a scope for the future work. 

%But we have not provided a proof.
 %It is also not clear whether first implies the second.   
\section*{Acknowledgments} It is pleasure to thank Sudipta Mukherji, Anirban Basu, Ajit Srivastava, Yogesh Srivastava, Shamik Banerjee for discussions and their comments. We are also thankful to other members of the string group of IOP for encouragement. 
  
\renewcommand{\thesection}{\Alph{section}}
\setcounter{section}{0}
\section{Expression for conductivity and flow equation} The electrical conductivity is usually computed from current-current correlator
\begin{eqnarray}
 \lambdabf &=& - \lim_{\omega \rightarrow 0} \frac{G_{xx}(\omega, q=0)}{i \omega}\nonumber\\
  &=& \lim_{\omega \rightarrow 0}\frac{1}{2 \omega}\int_{- \infty}^{\infty}dt~~e^{-i \omega t}\int d\vec{x}\langle[J_{x}(t,\vec{x}), J_{x}(0,\vec{0})]\rangle .\end{eqnarray} The current-current correlator can be computed by taking second derivative of effective action which reproduces the Eq.(\ref{eqnmotion1}) with respect to boundary fields.
Let us note that at low frequency, the  electrical conductivity (see \cite{Jain:2010ip} for details), at any radius $r$ can be written as
\begin{eqnarray}
\lambdabf(r) &=& -\frac{1}{\omega}\Bigg(\frac{g_{tt}}{g_{xx}}\Bigg)_{r} \Bigg(\frac{\rho^{2}}{\epsilon + P}\Bigg)_{r\rightarrow \infty} + \frac{1}{ g^{2}_{eff}}~ g_{xx}^\frac{d-3}{2}\Big{|}_{r=r_h} \frac{(sT)^{2}}{(\epsilon+ P)^{2}}\Big{|}_{r=r_c}\\
&=& -\frac{1}{\omega}\Bigg(\frac{g_{tt}}{g_{xx}}\Bigg)_{r} \Bigg(\frac{\rho^{2}}{\epsilon + P}\Bigg)_{r\rightarrow \infty} + \frac{1}{ g^{2}_{eff}}~ g_{xx}^\frac{d-3}{2}\Big{|}_{r=r_h} \frac{(s T_{H})^{2}}{(s T_{H} + \rho (A_{t}(r) - A_{t}(r_{h})))^{2}} \nonumber.
\end{eqnarray}
We also know how to relate the boundary transport coefficient with the horizon transport coefficient (see Eq (\ref{HBc})). So one of the remaining motivations to study flow equation can be to compute electrical conductivity away from low frequency limit (see \cite{Banerjee:2010zd}).
Effective action which will reproduce Eq (\ref{eqnmotion}) can be written as
\begin{eqnarray}
S &=& \frac {1}{2 \kappa^{2}} \int \frac{d^{d}q}{(2\pi)^d}dr \Big[-\frac{1}{2} N(r)\frac{d}{dr}\phi(r,\omega)\frac{d}{dr}\phi(r,-\omega)\nonumber \\
&+&\frac{1}{2} M(r)\phi(r,\omega)\phi(r,-\omega) -\omega^{2} \frac{1}{2}N(r)g_{rr} g^{tt}  \phi(r,\omega)\phi(r,-\omega)\Big],
\end{eqnarray}where
\begin{equation}
N(r)=\sqrt{-g}\frac{1}{g_{\rm{eff}}^2}g^{xx}g^{rr}
\end{equation}
and
\begin{equation}
 M(r)= \Big(\frac{1}{g_{\rm{eff}}^2}\Big)^2 \sqrt{-g} g^{xx}g^{rr}g^{tt}F_{rt}F_{rt}.
\end{equation}
The canonical momentum \footnote{Since $\phi = A_{x}$, momentum $\Pi \equiv J^{x}$,  where  $J^{x}$ is the current corresponding  to $A_{x}$ fluctuation.} conjugate to field $\phi(r,\omega)$ can be  written as,
\begin{eqnarray}
 \Pi(r,\omega) &=& \delta S \over \delta \phi'(r,\omega)\nonumber\\
&=& -\frac{1}{2 \kappa^{2}} N(r) \phi' ,
\end{eqnarray} where $\phi' = \frac{d}{dr} \phi $. Now we can define electrical conductivity as
\begin{equation}
 \sigma(r,\omega) = \frac{\Pi(r,\omega)}{i \omega \phi(r)}.
\end{equation}
If we define $\sigma(r,\omega)= i~ \Im(\sigma(r,\omega)) + \Re(\sigma(r,\omega)) $, then we get
\begin{eqnarray}
 \Re(\sigma(r,\omega))  &=& \Re\Bigg(\frac{\Pi(r,\omega)}{i \omega \phi(r)}\Bigg)\nonumber\\
&=& \Re\Bigg(\frac{\Pi(r,\omega)\phi(r)}{i \omega \phi^2(r)}\Bigg)\nonumber\\
&=& -\Im\Bigg(\frac{\Pi(r,\omega)\phi(r)}{ \omega \phi^2(r)}\Bigg).
\end{eqnarray}
Using the fact that 
\begin{equation}
 \frac{d}{dr}\Im[\Pi(r,\omega)\phi(r)] =0,
\end{equation} and 
\begin{equation}
\lim_{r\rightarrow r_{h}}\frac{d}{dr}\phi(r)=-i \omega \lim_{r\to r_{h}}\sqrt{\frac{g_{rr}}{g_{tt}}} \phi(r)+{\mathcal{O}}(\omega^{2}). 
\end{equation}
we get (in the limit $\omega \to 0$)

\begin{eqnarray}
\Re(\sigma(r)) &=& \frac {1}{2\kappa^{2}} \Bigg(\sqrt{\frac{g_{rr}}{g_{tt}}}  N(r)\Bigg)_{r=r_{h}} \Bigg(\frac{\phi(r_{h})}{\phi(r)}\Bigg)^{2} \nonumber\\
 &=& \frac {1}{2 \kappa^{2}}\Bigg( \frac{1}{g_{\rm{eff}}^2}~ g_{xx}^\frac{d-3}{2} \Bigg)_{r=r_{h}}\Bigg(\frac{\phi(r_{h})}{\phi(r)}\Bigg)^{2}\nonumber\\
&=& \sigma_{H}~~\Bigg(\frac{\phi(r_{h})}{\phi(r)}\Bigg)^{2} ,\label{sincleconduc11}
\end{eqnarray} 
where $\sigma_{H} $ is the conductivity evaluated at the horizon and its expression is given by,
\begin{equation}
\sigma_{H} = \frac {1}{2 \kappa^{2}g_{\rm{eff}}^2}~ g_{xx}^\frac{d-3}{2}\Big{|}_{r=r_{h}}.\end{equation}At the boundary we get
\begin{equation}
\Re(\sigma(r\rightarrow \infty)) = \sigma_{H}~~\Bigg(\frac{\phi(r_{h})}{\phi(r\rightarrow \infty)}\Bigg)^{2}.\end{equation} The imaginary part of conductivity can be written as
\begin{equation}
 \Im(\sigma(r\rightarrow \infty))= -\frac{1}{\omega}\lim_{r\to \infty}\lim_{\omega \to 0}\frac{\Pi(r,\omega)}{\phi(r,\omega)}
\end{equation} We refer reader to \cite{Jain:2010ip} for details and regarding how to compute electrical conductivity for multiple charge case.
In the following we turn our attention to the case away from low frequency limit.
Taking derivative with respect to $r$, we get
\begin{equation}
 \frac{d}{dr} \sigma(r,\omega) = i \omega \frac{2\kappa^{2}}{N(r)} \big[\sigma^{2}(r,\omega) + (\frac{1}{2 \kappa^{2}})^{2} (N^{2}(r) g_{rr} g^{tt} - \frac{1}{\omega^{2}}M(r) N(r))\big],
\end{equation} where we have used Eq (\ref{eqnmotion}) and 
\begin{equation}
 \frac{d}{dr} \Pi(r,\omega) = -\frac{\omega^{2}}{2 \kappa^{2}}(N(r) g_{rr} g^{tt}\phi(r) - \frac{1}{\omega^{2}}M(r)\phi(r)).
\end{equation}
If we define $\sigma(r,\omega)= i~ \Im(\sigma(r,\omega)) + \Re(\sigma(r,\omega)) $, then we get
\begin{equation}
 \frac{d}{dr} \Re(\sigma(r,\omega)) = -\omega \frac{4\kappa^{2}}{N(r)} \Re(\sigma(r,\omega)) \Im(\sigma(r,\omega)),
\end{equation}
\begin{equation}
\frac{d}{dr} \Im(\sigma(r,\omega)) = \omega \frac{2\kappa^{2}}{N(r)} \big[\Big(\Re(\sigma(r,\omega))\Big)^{2} - \Big(\Im(\sigma(r,\omega))\Big)^{2} + (\frac{1}{2 \kappa^{2}})^{2} (N^{2}(r) g_{rr} g^{tt} - \frac{1}{\omega^{2}}M(r) N(r))\big].
\end{equation}
By solving above equations perturbatively in $\omega$ or numerically we can get electrical conductivity away from low frequency limit. Let us note that,
if we take spatial momentum also to be non zero then gauge field fluctuation and metric fluctuation no longer decouple. In this case computation may become much more subtle. 

\section{Condition on energy momentum tensor}
Let us consider a constant $r$ hypersurface outside the horizon. The unit normal vector to that hypersurface is $n^{\mu}\d_{\mu} = n^{r}\d_{r}$, where $n^{r} = \sqrt{g^{rr}}.$ One can define the extrinsic curvature $\Theta_{\mu\nu}$ of the hypersurface to be
\begin{equation}
 \Theta_{\mu\nu}= -\frac{1}{2}(\bigtriangledown_{\mu} n_{\nu} + \bigtriangledown_{\nu} n_{\mu}).
\end{equation}
Using the form of the metric as in Eq (\ref{met1}), we get
\begin{equation}
 \Theta_{tt} = -\frac{1}{2}\sqrt{g^{rr}} \frac{d}{dr}g_{tt}~~, ~~~~~~~ \Theta_{xx} = -\frac{1}{2}\sqrt{g^{rr}} \frac{d}{dr}g_{xx}.
\end{equation}
Using Eq (\ref{rxx}) and Eq (\ref{rtt}), we can write
\begin{equation}
 \sqrt{g} R^{t}_{t} = \frac{d}{dr}(\sqrt{h} \Theta^{t}_{t}),~~~ \sqrt{g} R^{x}_{x} = \frac{d}{dr}(\sqrt{h} \Theta^{x}_{x}),
\end{equation} where $h$ is the determinant of the induced metric on the hypersurface. The induced metric on the constant $r$ hypersurface is given by
\begin{eqnarray}
 ds^{2}_{\sum } &=& h_{tt}dt^{2} + h_{xx} \sum_{i=1}^{d-1} (dx^{i})^{2}\nonumber\\
&=& g_{tt}dt^{2} + g_{xx} \sum_{i=1}^{d-1} (dx^{i})^{2}\label{meth}.
\end{eqnarray} Let us define a tangent null vector
$l^{\mu}\d_{\mu} = \sqrt{-g^{tt}}\d_{t} +  \sqrt{g^{xx}}\d_{x}$.
 Now we can write Eq (\ref{rxt}) and consequently Eq (\ref{pc}) as
\begin{eqnarray}
 \sqrt{-g} R_{\mu\nu}l^{\mu}l^{\nu} &=& \sqrt{-g} T^{Total}_{\mu\nu}l^{\mu}l^{\nu}\nonumber\\
 &=& \sqrt{-g}T^{E.M.}_{\mu\nu}l^{\mu}l^{\nu}+\sqrt{-g}T^{Matter}_{\mu\nu}l^{\mu}l^{\nu},
\end{eqnarray}
\begin{eqnarray}
 \sqrt{-h}\Theta_{\mu\nu}l^{\mu}l^{\nu}\Big|_{r_{h}}^{r} &=&  \int_{r_{h}}^{r} dr \sqrt{-g}T^{E.M.}_{\mu\nu}l^{\mu}l^{\nu}+ \int_{r_{h}}^{r} dr \sqrt{-g}T^{Matter}_{\mu\nu}l^{\mu}l^{\nu}\nonumber\\
&=&-\kappa^{2} \rho A_t\Big{|}_{r_{h}}^{r}+\int_{r_{h}}^{r} dr \sqrt{-g}T^{Matter}_{\mu\nu}l^{\mu}l^{\nu},
\end{eqnarray} respectively. Upon using the Einstein equation (\ref{ein}) and the fact that for the metric of the form given in Eq (\ref{met1}), the $R_{xt}$ component of the Ricci tensor is zero, we get $T^{Matter}_{tx} =0,$ since $T^{E.M.}_{tx}=0$. So the condition that we get on the energy momentum tensor\footnote{According to null energy condition, $T^{total}_{\mu\nu}l^{\mu}l^{\nu}\geq 0,$ with $l^{\mu}$ a null vector. Since $T^{E.M}l^{\mu}l^{\nu} \geq 0$,  the contribution from the matter part $T^{Matter}_{\mu\nu}l^{\mu}l^{\nu}$ may be negative as well. However it is interesting to note that, if we take a limit where charge of the black hole vanishes then $T^{E.M}l^{\mu}l^{\nu} = 0$, so that null energy condition gives $T^{Matter}_{\mu\nu}l^{\mu}l^{\nu}\geq 0$.  So if we are interested in the backgrounds where matter sector does not act as source for electromagnetic field, it appears that $T^{Matter}_{\mu\nu}l^{\mu}l^{\nu}\geq 0,$ irrespective of presence of gauge fields.} in Eq (\ref{condition}) can be written as
\begin{equation}
 T^{Matter}_{\mu\nu}l^{\mu}l^{\nu} =0.
\end{equation}

\end{document}